# Registration-based Compensation using Sparse Representation in Conformal-array STAP

Ke Sun[1], Huadong Meng[1], Fabian Lapierre[2], Xiqin Wang[1]

([1]Department of Electronic Engineering, Tsinghua University, Beijing 100084, China

[2]Royal Military Academy, Department of Electrical Engineering, Brussels 1000, Belgium)

**Abstract**: Space-time adaptive processing (STAP) is a well-suited technique to detect slow-moving targets in the presence of a clutter-spreading environment. When considering the STAP system deployed with conformal radar array (CFA), the training data is range-dependent, which results in poor detection performance of traditional statistical-based algorithms. Current registration-based compensation (RBC) is implemented based on sub-snapshot spectrum using temporal smoothing. In this case, the estimation accuracy of the configuration parameters and the clutter power distribution is limited. In this paper, we introduce the technique of sparse representation into the spectral estimation and propose a new compensation method, called RBC with sparse representation (SR-RBC). This method first converts the clutter spectral estimation into an ill-posed problem with the constraint of sparsity. Then the technique of sparse representation like iterative reweighted least squares (IRLS) is utilized to solve this problem. Based on this, the transform matrix is designed so that the processed training data behaves nearly stationary with the test cell. Since the configuration parameters as well as the clutter spectral response are obtained with full-snapshot using sparse representation, SR-RBC provides more accurate clutter spectral estimation and the transformed training data is more stationary so that better signal-clutter-ratio (SCR) improvement is expected.

# 1. Introduction

An airborne/spaceborne space-time adaptive processing (STAP) is the technique of choice to detect slow-moving targets in the presence of a strong clutter background. Conventional STAP processors using a side-looking uniform linear array (ULA) have the desirable property that the relationship between the clutter spatial and Doppler frequencies is range-independent. Thus the training data from adjacent range cells behaves stationary and can be utilized to estimate the clutter covariance matrix (CCM) so that the adaptive filter can be effectively contracted to improve the output signal-clutter-ratio (SCR) in the test cell [1-2]. However in many radar and sonar applications, achieving perfectly ULA geometries may not always be practical. Besides, complex configuration, e.g., conformal radar array (CFA) does provide certain advantages including minimal payload weight, the potential for increased aperture, and increased field of view [3-4]. However, the relationship between the clutter spatial-Doppler frequencies becomes nonlinear and range-dependent at this case of CFA. Thus the sample covariance matrix computed from the training data set mismatches with the test cell and results in degraded performance in canceling clutter.

To deal with the range-dependent clutter, various methods have been proposed. Angle-Doppler compensation (ADC) and adaptive angle-Doppler compensation (A2DC) [5-6] attempt to align the peaks of the clutter ridge of the training data with that of the test cell. However, they only accomplish partial compensation, and thus are effective only for highly directive antenna beampatterns. Derivative-based updating (DBU) method assumes that the clutter central Doppler frequency is linear with range. However, this assumption is rarely satisfied at the short range case and thus DBU also does not work effectively with all configurations such as CFA and/or bistatic radar [7]. Recently, the registration-based compensation (RBC) [8-10] method is proposed thorough a mathematical description of the clutter ridge in the angle-Doppler domain. This method estimates both the configuration parameters and the clutter power distribution. Based on this, the transform matrix is designed so that the training data is nearly stationary with the test cell. RBC can implement both the mainlobe and sidelobe clutter compensation. However, since the peak extraction is implemented in the sub-snapshot spectrum, the estimation accuracy of both the configuration parameters and clutter power distribution is limited, which results in degraded performance. In this paper, we introduce the technique of sparse representation into the problem of clutter spectral estimation and propose a novel registration-based compensation called SR-RBC to further improve the SCR performance. The remainder of this paper is organized as follows. Section 2 describes the basic signal model deployed with the CFA configuration. Section 3 introduces the theory of the sparse representation and illustrates the details of the SR-RBC algorithm. Section 4 uses the simulated data to illustrate the advantages of the proposed method. Section 5 gives a conclusion of the proposed method and points out the future work.

## 2. CFA Signal Model

Conformal antenna assumes the shape of the radar-bearing platform and generally belongs to the class of nonlinear array. Specific advantages of conformal antenna include better aerodynamic shape compatible with the airframe, potentially greater effective apertures, less payload weight and so on. Thus, STAP deployed with conformal antenna has great potential in the future airborne radar system, especially, unmanned aerial vehicle [3-4]. In this section, we discuss the space–time response of the conformal array to the ground clutter scatters. Suppose the conformal antenna is deployed onto a particular surface of the radar platform– such as the fuselage, wing or nose cone

–the array response is generally non-linear from element to element comprising the multichannel receive array. To determine the response of a point clutter scatter, we rely on the coordinate system of Fig.1, with the x-axis aligned to true north, the y-axis pointing to west and the z-axis perpendicularly directed away from the Earth's surface. Here we adopt the cylindrical arrays with $M$ rings, each of which is composed of $N$ isometry array elements. The parallel rings are perpendicular to the y-axis at a spacing of $d$. The array elements within each ring are isometry placed with a circle radius $r$. Although the discussion in this paper is carried out with the cylindrical array, our method can be effectively implemented in other CFA configurations. In this instance, we defines angle vector $\mathbf{\psi} = [\phi, \theta]^T$, where symbols $\phi, \theta$ stand for the azimuth and elevation angles of certain clutter scatter $Q$, respectively.

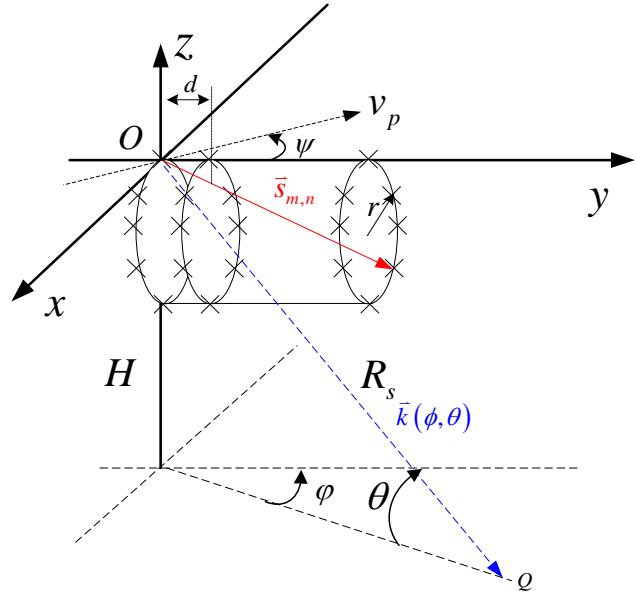

Fig.1 Geometry of the CFA configuration

The unit vector $\vec{\mathbf{k}}(\mathbf{\psi})$ points orthogonal to the propagating planar wavefront and is given as

$$\vec{\mathbf{k}}(\mathbf{\psi}) = \cos\theta\cos\phi \cdot \vec{e}_x + \cos\theta\sin\phi \cdot \vec{e}_y - \sin\theta \cdot \vec{e}_z. \tag{1}$$

Additionally, the direction vector to the $nth$ element on the $mth$ ring is defined as

$$\bar{\mathbf{s}}_{m,n} = r \cdot \sin\left[\frac{2\pi}{N}(n-1)\right] \cdot \bar{e}_x + d(m-1) \cdot \bar{e}_y - r \cdot \cos\left[\frac{2\pi}{N}(n-1)\right] \cdot \bar{e}_z. \quad (2)$$

From [1], the element-level spatial response at the $kth$ range cell is then given as

$$\begin{aligned}\mathbf{x}_{s/k} &= \varsigma v_k \left[ a_{s,1,1}\sqrt{g_{1,1}} e^{j(2\pi/\lambda)\bar{\mathbf{k}}(\psi)\cdot\bar{\mathbf{s}}_{1,1}}, \cdots a_{s,1,N}\sqrt{g_{1,N}} e^{j(2\pi/\lambda)\bar{\mathbf{k}}(\psi)\cdot\bar{\mathbf{s}}_{1,N}} \right. \\ &\quad \left. , \cdots a_{s,M,N}\sqrt{g_{M,N}} e^{j(2\pi/\lambda)\bar{\mathbf{k}}(\psi)\cdot\bar{\mathbf{s}}_{M,N}} \right]^T \\ &= \varsigma v_k \left[ \mathbf{a}_s \odot \mathbf{g} \odot \mathbf{s}_s(\psi) \right],\end{aligned} \quad (3)$$

where $\varsigma \sim CN(0,1)$ (if the response is non-fluctuating, then $\varsigma$ is a complex scalar with unity magnitude and uniformly distributed phase), $v_k$ is a normalized voltage term following from the radar range equation. The spatial covariance matrix taper (CMT) $\mathbf{a}_s$ reflects the inter-array amplitude-phase inconsistency with $E\left[\mathbf{a}_t\mathbf{a}_t^H\right] = \mathbf{A}_t$. $\mathbf{g} = \left[\sqrt{g_{1,1}}, \cdots \sqrt{g_{1,N}}, \cdots \sqrt{g_{M,N}}\right]^T$ is the antenna gain vector, and $\odot$ is the Hadamard product. Thus the space steering vector is defined as

$$\mathbf{s}_s(\psi) = \left[ e^{j(2\pi/\lambda)\bar{\mathbf{k}}(\psi)\cdot\bar{\mathbf{s}}_{1,1}}, \cdots e^{j(2\pi/\lambda)\bar{\mathbf{k}}(\psi)\cdot\bar{\mathbf{s}}_{1,N}}, e^{j(2\pi/\lambda)\bar{\mathbf{k}}(\psi)\cdot\bar{\mathbf{s}}_{M,N}} \right]^T. \quad (4)$$

Simultaneously, the Doppler steering vector describes the pulse-to-pulse phase change due to the platform and target motion. Considering a stationary clutter scatter $Q$ at angle $\psi$, the corresponding Doppler steering vector is given as

$$\mathbf{s}_t(\psi) = \left[ e^{j2\pi(2\bar{k}(\psi)\cdot\bar{\mathbf{v}}_p/\lambda)T}, e^{j2\pi(2\bar{k}(\psi)\cdot\bar{\mathbf{v}}_p/\lambda)2T}, \ldots, e^{j2\pi(2\bar{k}(\psi)\cdot\bar{\mathbf{v}}_p/\lambda)(P-1)T} \right]^T, \quad (5)$$

where $\bar{\mathbf{v}}_p$ represents the platform velocity vector and is given as

$$\bar{\mathbf{v}}_p = -|\bar{\mathbf{v}}_p|\sin\sigma \cdot \bar{e}_x + |\bar{\mathbf{v}}_p|\cos\sigma \cdot \bar{e}_y, \quad (6)$$

where $\sigma$ denotes the crab angle between the flight direction and the central axis of the cylindrical array. Symbol $T$ is the pulse repetition interval and $P$ is the total number of pulses comprising the temporal aperture. A simple modification to (5) is needed if the scatter is moving, i.e., the moving target. Based on this, the temporal snapshot of certain scatter $Q$ at the $(m,n)th$ array element, the $kth$ range cell is given as

$$\mathbf{x}_{t/k}(m,n) = \varsigma v_k \sqrt{g_{m,n}} \left( \mathbf{a}_t \odot \mathbf{s}_t(\psi) \right), \quad (7)$$

where $\mathbf{a}_t$ is a temporal CMT with $E[\mathbf{a}_t \mathbf{a}_t^H] = \mathbf{A}_t$, which is caused by the intrinsic clutter motion and system jitter etc. Both spatial and temporal decorrelation expands the clutter rank and spreads the clutter ridge in the angle-Doppler domain. Thus, the space–time response of the CFA to a stationary scatter at angle $\psi$ conveniently follows as

$$\mathbf{x}_k = \varsigma v_k \left(\mathbf{a}_t \odot \mathbf{s}_t(\psi)\right) \otimes \left(\mathbf{a}_s \odot \mathbf{g} \odot \mathbf{s}_s(\psi)\right), \tag{8}$$

where $\otimes$ indicates the Kronecker product. Space–time correlation taper $\mathbf{a}_{s-t} = \mathbf{a}_s \otimes \mathbf{a}_t$ satisfies $\mathbf{A}_{s-t} = E[\mathbf{a}_{s-t} \mathbf{a}_{s-t}^H] = \mathbf{A}_t \otimes \mathbf{A}_s$. A realistic model for the ground clutter return results from the coherent summation of many clutter scatterings within the bounds of each iso-range [4]. Thus the clutter space–time snapshot at the $kth$ range cell takes the form as

$$\mathbf{c}_k = \sum_{p=1}^{N_a} \sum_{q=1}^{N_c} \varsigma(\psi_{p,q}, k) v(\psi_{p,q}, k) \left(\mathbf{a}_t \odot \mathbf{s}_t(\psi_{p,q}, k)\right) \\ \otimes \left(\mathbf{a}_s \odot \mathbf{g}(\psi_{p,q}, k) \odot \mathbf{s}_s(\psi_{p,q}, k)\right), \tag{9}$$

where $N_a$ indicates the number of ambiguous range cells, $N_c$ is the number of statistically independent clutter scatters at each iso-range, $\psi_{p,q}$ indicates the certain angle vector of the $qth$ clutter scatter at the $pth$ iso-range. Moreover, by virtue of statistical independence of each clutter scatter, the clutter space–time covariance matrix of the $kth$ range cell follows as [4]

$$\mathbf{R}_k = \sum_{p=1}^{N_a} \sum_{q=1}^{N_c} \sigma_s^2(\psi_{p,q}, k) \mathbf{A}_{s-t}(\psi_{p,q}, k) \odot \mathbf{g}_{s-t}(\psi_{p,q}, k) \mathbf{g}_{s-t}^H(\psi_{p,q}, k) \\ \odot \mathbf{s}_{s-t}(\psi_{p,q}, k) \mathbf{s}_{s-t}^H(\psi_{p,q}, k). \tag{10}$$

In this case, the clutter is substantially range-dependent so that traditional CCM estimation such as loaded sample matrix inversion (LSMI) tends to behave the average behavior of the training data. Thus the corresponding STAP filter response exhibits mismatch for the test cell, with insufficient null depth and excessive clutter spread, which causes performance degradation.

A series of methods have been proposed to deal with the range-dependent clutter. Methods such as angle-Doppler compensation and adaptive angle-Doppler compensation [5-6] accomplish the peak response, but the sidelobe clutter suppression is limited. RBC implements both the mainlobe and sidelobe compensation [8-10]. However, since the configuration parameters as well as the clutter power distribution are obtained using the sub-snapshot spectrum, the estimation is not accurate enough. Thus, the stationarity of the processed training data is destroyed and causes performance degradation of the STAP filter. In the following part, we propose to make range-dependent compensation using the technique of sparse representation, which has the capability of obtaining more accurate clutter response in the full-snapshot spectrum so that the compensation performance is further improved.

## 3. Range-Dependent Compensation using Sparse Representation

The key requirement for STAP with any geometry configuration is the accurate knowledge of the clutter spectral response (i.e., the shape of the clutter ridge in the angle-Doppler domain) [1-2]. In the common case of ULA, the training data behaves stationary and can be utilized to estimate the accurate clutter response (termed as CCM) of the test cell. In the non side-looking and/or CFA STAP cases, the clutter behaves range-dependent. To solve this problem, a series of preprocessings are proposed. RBC utilizes the technique of temporal smoothing to obtain sub-snapshot spectrum and then generates the transform matrix at each range cell so that the processed training data behaves nearly stationary. Thus, the sub-snapshot spectrum is the key to guarantee the desirable performance in the subsequent processings such as the estimation of configuration parameters and clutter power distribution. In this paper, we also seek to require accurate clutter spectral response, which is similar to that in RBC. However, the difference lies in that SR-RBC can obtain more accurate clutter spectrum with full-snapshot and the estimation of the configuration parameters is avoided. In the following part, the technique of sparse representation is first introduced and then utilized into our problem of the clutter spectral estimation. Based on this, the procedures of the overall algorithm are elaborated.

### 3.1 Clutter spectral estimation using sparse representation

First discretize the angle and Doppler frequency axes into $N_s = \rho_s NM, N_t = \rho_t P$ grids in the angle-Doppler domain. The parameters $\rho_s, \rho_t$ are the zoom scales along the angle and Doppler axes, respectively. Let $\phi_i = \frac{2\pi i}{N_s}, 1 \le i \le N_s$ and $f_{d,j} = \frac{j}{N_d}, 1 \le j \le N_d$ denote the uniformly-discretized azimuth angles and Doppler frequencies, respectively. The corresponding angle vectors for the $kth$ range cell are given as

$$\mathbf{\psi}_{i,k} = [\phi_i, \theta_k]^T, 1 \le i \le N_s \tag{11}$$

Based on this, the received data of the $kth$ range cell can be written in matrix form as

$$\mathbf{x}_k = \sum_{i=1}^{N_s N_t} \alpha_i \cdot \mathbf{\Phi}_{i,k} + \mathbf{n}_k = \mathbf{\Phi}_k \mathbf{\alpha}_k + \mathbf{n}_k, \tag{12}$$

where the $NMP \times N_s N_t$ matrix $\mathbf{\Phi}_k$ is the overcomplete basis composed with all the possible space-time steering vectors as

$$\mathbf{\Phi}_k = \left[ \mathbf{s}_{s-t}(\mathbf{\psi}_{1,k}, f_{d,1}), \cdots, \mathbf{s}_{s-t}(\mathbf{\psi}_{N_s,k}, f_{d,1}), \cdots, \mathbf{s}_{s-t}(\mathbf{\psi}_{N_s,k}, f_{d,N_d}) \right]. \tag{13}$$

The vector $\mathbf{\alpha}_k$ stands for the spectral distribution of the $kth$ range cell in the basis $kth$, (i.e., the space-time spectrum in the angle-Doppler domain), and $\mathbf{n}_k$ is the observation noise. Equation (12) is the fundamental equation in this paper and has two characteristics that we should pay attention to. First, estimating the space-time spectrum $\mathbf{\alpha}_k$ is equivalent to solving the linear equation (12) with the data $\mathbf{x}_k$. Second, the basis $\mathbf{\Phi}_k$ is overcomplete and the problem is ill-posed because the zoom scales $\rho_s, \rho_t$ are greater than one to obtain the high-resolution spectrum. Generally, when the positions of the actual clutter scatters are known in advance, this ill-posed problem can be simplified into an overdetermined equation, which can be effectively solved by least squares (LS) [6]. However, this prior knowledge is hard to guarantee in the actual clutter scenario. On the other hand, the theory of sparse recovery has proved that: even when the actual positions are unknown, the ill-posed problem can be effectively solved provided that the actual clutter spectral distribution $\boldsymbol{\alpha}_0$ is sparse [11-12]. Next the sparsity of the clutter spectral response is first illustrated.

As shown in Fig.2, after the discretization, each cell in this plane corresponds to a certain space-time steering vector and all of these vectors comprise the overcomplete basis $\boldsymbol{\Phi}_k$. Since the STAP clutter scenario usually has a high CNR [1-2], the distribution in the angle-Doppler plane is mainly determined by the clutter distribution. Due to the angle-Doppler dependence of the clutter scatters, the significant elements of the spectral distribution only focuses along the clutter ridge in the angle-Doppler domain, whose slope is determined by the radar configuration parameters and behaves range-dependent in the case of CFA. Thus the clutter spectrum is sparse, i.e., only a small amount of elements are significant and others are quite small. This statement is even valid in the case of omnidirectional antenna, where the clutter scatters come from all the directions but the cells occupied by the clutter ridge is still small compared with the whole angle-Doppler plane.

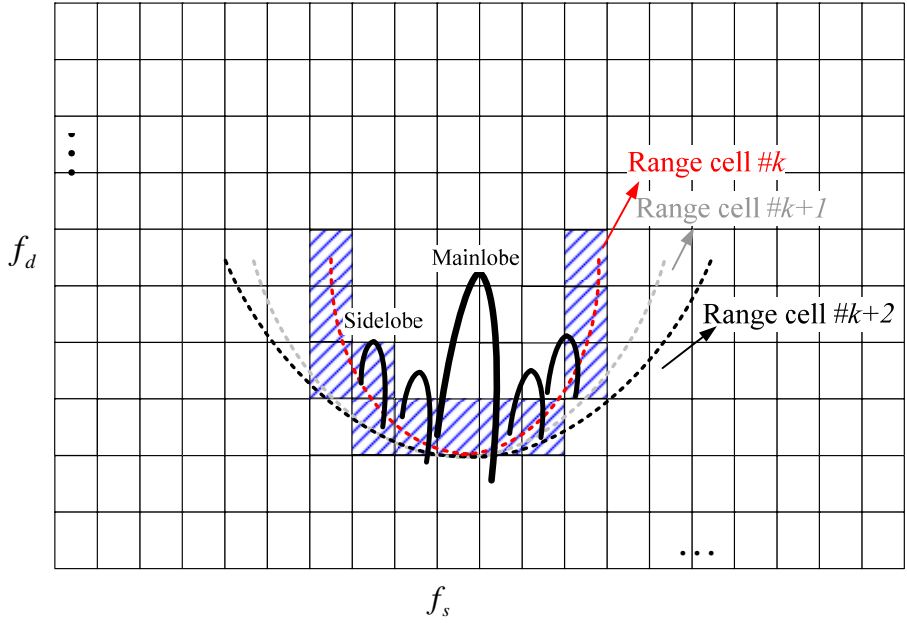

Fig.2 clutter spectral response of CFA STAP

## 3.2 Sparse-induced Compensation for range-dependent clutter

According to the theory of sparse representation [11-12], when the actual distribution is sparse in a domain, the ill-posed problem in (12) can be efficiently solved. The basic form of sparse representation is defined as

$$\hat{\boldsymbol{\alpha}} = \arg\min \|\boldsymbol{\alpha}\|_0 \ \ subject\ to\ \ \|\mathbf{x} - \boldsymbol{\Psi}\boldsymbol{\alpha}\|_2 \leq \varepsilon, \qquad (14)$$

where $\varepsilon$ is the data fitting allowance, $\varepsilon$ stands for the $L_p$ norm and thus $\|\cdot\|_0$ denotes the number of the nonzero elements of a vector. However, this optimization is a combinatorial problem and NP-hard. To address this difficulty, a number of practical algorithms have been proposed to approximate this sparse solution. One way is to replace the objective function with the $L_1$ norm [12]. It has been proven that this approximation can achieve quite desirable performance but demands a high computational effort especially for large-scale problems. Besides, a series of fast and greedy approximations are proposed, among which iterative reweighted least square (IRLS) appears to be both effective and time-saving [13-15]. IRLS uses the reweighted $L_2$ norm minimization to make recursive adjustments to the weightings until most of the elements in the solution are close to zero and generate a sparse solution. IRLS has been widely used in the applications such as source localization and neuromagnetic imaging [16-17], however, the potential seems to be more than current applications.

In this paper, we propose a novel method called SR-RBC to make registration-based compensation (the initial idea is developed from our earlier work [18]). This method has the advantage of serving two purposes simultaneously. First, the estimation of the configuration parameters and the clutter power distribution is integrated into the solution of ill-posed problem with the constraint of sparsity so that the estimation is data-based and no prior knowledge is needed. Second, the sparse representation such as IRLS can effectively solve the estimation problem and obtain more accurate clutter spectral response with full-snapshot. Thus the transformed training data behaves more stationary with the test cell so that the SCR performance is further improved. The details of the whole algorithm are given as follows.

### 3.2.1 Clutter spectral estimation using IRLS

Since IRLS is served as an iterative algorithm, it only guarantees a local sparse solution, which may not coincide with the actual solution. Thus, appropriate initial value is necessary to assure the final convergence. Here we adopt the Fourier spectrum as

$$\alpha_{i,k}^{(0)} = \mathbf{\Phi}_{i,k}^{H}\mathbf{x}_{k}, 1 \leq i \leq N_{s}N_{t}, \tag{15}$$

where the initial vector $\mathbf{\alpha}_{k}^{(0)}$ is unbiased but low-resolution. Although current strategy is adopted as the Fourier spectrum, the analysis in [13-14] has illustrated that any unbiased initialization is also effective. Furthermore, the initialization does not have to be sparse, otherwise, some potentially elements may be lost and not recovered in the subsequent iterations. Additionally, the initial weighting matrix and the adaptive subspace are given as

$$\mathbf{W}^{(0)} = \text{diag}\left(\left[\text{abs}\left(\boldsymbol{\alpha}_k^{(0)}\right)\right]\right),$$
$$\Gamma = \{i, 1 \leq i \leq N_s N_t\},$$
(16)

where $\text{diag}(\cdot)$ is the diagonalization. Then the estimation updating at the $lth$ iteration is given accordingly as

$$\boldsymbol{\alpha}_k^{(l)}\big|_\Gamma = \mathbf{W}^{(l)} \left(\boldsymbol{\Phi}_k\big|_\Gamma \mathbf{W}^{(l)}\right)^\dagger \mathbf{x}_k,$$
(17)

where $\boldsymbol{\alpha}_k^{(l)}\big|_\Gamma$ stands for the $\Gamma$ subset of the vector $\boldsymbol{\alpha}_k^{(l)}$, $(\mathbf{A})^\dagger = \left(\mathbf{A}^H \mathbf{A}\right)^{-1} \mathbf{A}^H$ denotes the pseudoinverse operation of matrix $\mathbf{A}$. As the iterations is implemented, some of the elements in the estimation $\boldsymbol{\alpha}_k^{(l)}$ become close to zero, thus the procedure of the reweighted least square in (17) can be only carried on a subspace $\boldsymbol{\Phi}_k\big|_\Gamma$. Additionally, the dimension of the subspace should also be adjusted during the iteration as

$$\Gamma = \arg\left(\left|\alpha_{i,k}^{(l)}\right| \geq Th\right), \ 1 \leq i \leq N_s N_t,$$
(18)

where $Th$ stands for the threshold and $\alpha_{i,k}^{(l)}$ denotes the $ith$ element of the solution $\boldsymbol{\alpha}_k^{(l)}$. Based on this, the weighting matrix can be updated as

$$\mathbf{W}^{(l+1)} = \text{diag}\left(\left\{\alpha_{i,k}^{(l)}, i \in \Gamma\right\}\right).$$
(19)

Finally, the convergence judgment is made as

$$\left|\frac{\boldsymbol{\alpha}_k^{(l)} - \boldsymbol{\alpha}_k^{(l-1)}}{\boldsymbol{\alpha}_k^{(l)}}\right| \leq \varsigma,$$
(20)

where $\varsigma$ stands for a small constant. Otherwise, repeat the recursive process as (17)-(19). After obtaining the high-resolution clutter spectral estimation, the CCM estimation can be given as

$$\hat{\mathbf{R}}_{c,k} = \sum_{i \in \Gamma} \left|\hat{\alpha}_k(i)\right|^2 \boldsymbol{\Phi}_{i,k} \boldsymbol{\Phi}_{i,k}^H + \beta_L \mathbf{I},$$
(21)

where $\hat{\alpha}_k(i)$ is the $ith$ element of the final spectral estimation $\hat{\boldsymbol{\alpha}}_k$ at the $kth$ range cell, $\beta_L$ is a small loading to match the noise level.

**3.2.2 Range-dependent compensation using SR-RBC**

The idea of designing the transform matrix using CCM so that the processed training data is stationary with the test cell is first proposed in RBC [9]. In this paper, SR-STAP follows the similar idea with RBC, but since the clutter spectrum and the corresponding CCM estimation is obtained with higher accuracy, the stationarity of the processed training data can be further improved. In other words, we seek to generate the matrix $\mathbf{T}_k$ so that the processed training data $\tilde{\mathbf{x}}_k = \mathbf{T}_k \mathbf{x}_k$ is stationary with the test cell $\mathbf{x}_t \sim CN(0, \mathbf{R}_{c,t})$. Substituting the CCM of the processed data into the stationarity definition, we can deduce that

$$\tilde{\mathbf{R}}_{c,k} = E\left[\tilde{\mathbf{x}}_k \tilde{\mathbf{x}}_k^H\right] = \mathbf{T}_k \mathbf{R}_{c,k} \mathbf{T}_k^H = \mathbf{R}_{c,t}. \tag{22}$$

Using the eigenvalues decomposition, $\mathbf{R}_{c,k}$ and $\mathbf{R}_{c,t}$ can be expressed as

$$\mathbf{R}_{c,k} = \mathbf{V}_k \mathbf{\Lambda}_k \mathbf{V}_k^H = \left(\mathbf{V}_k \mathbf{\Lambda}_k^{1/2}\right)\left(\mathbf{V}_k \mathbf{\Lambda}_k^{1/2}\right)^H, \tag{23}$$

$$\mathbf{R}_{c,t} = \left(\mathbf{V}_t \mathbf{\Lambda}_t^{1/2}\right)\left(\mathbf{V}_t \mathbf{\Lambda}_t^{1/2}\right)^H, \tag{24}$$

where $\mathbf{\Lambda}_k$ and $\mathbf{\Lambda}_t$ are the diagonal matrixes containing the eigenvalues for the $kth$ training data and the test cell, respectively, $\mathbf{V}_k$ and $\mathbf{V}_t$ are the corresponding matrixes containing the eigenvectors as columns. Based on this, the processed CCM at the $kth$ rang cell $\tilde{\mathbf{R}}_{c,k}$ is further expressed as

$$\tilde{\mathbf{R}}_{c,k} = \left(\mathbf{T}_k \mathbf{V}_k \mathbf{\Lambda}_k^{1/2}\right)\left(\mathbf{T}_k \mathbf{V}_k \mathbf{\Lambda}_k^{1/2}\right)^H. \tag{25}$$

Thus, to generate the stationary training data as $\tilde{\mathbf{R}}_{c,k} = \mathbf{R}_{c,t}$, we have

$$\mathbf{T}_k \mathbf{V}_k \mathbf{\Lambda}_k^{1/2} = \mathbf{V}_t \mathbf{\Lambda}_t^{1/2}, \tag{26}$$

then the transform matrix is correspondingly given as

$$\mathbf{T}_k = \mathbf{V}_t \mathbf{\Lambda}_t^{1/2} \mathbf{\Lambda}_k^{-1/2} \mathbf{V}_k^H. \tag{27}$$

In this way, the processed training data $\tilde{\mathbf{x}}_k$ behaves nearly stationary with the test cell and can be utilized to estimate the CCM of the test cell using statistical-based methods like LSMI.

To sum up, SR-RBC seeks to generate the transform matrix using the clutter spectral estimation, which follows the basic idea of RBC. However, there exists some critical difference between them, which is thought to be the reason of performance improvement. RBC generates the sub-snapshot spectral response in the angle-Doppler domain and selects peaks to estimate the configuration parameters using the curve-fitting. Based on this, the clutter power distribution and the CCM estimation is obtained at each range cell. On the contrary, SR-RBC combines the procedures of estimating both the configuration parameters and the clutter power distribution into the solution of ill-posed problem with the constraint of sparsity. Since the spectral estimation in SR-RBC is carried out with full-snapshot, no degree of freedom （DOF） loss is generated and more accurate spectral estimation is expected. Besides, since the prior knowledge is not required in the spectral estimation, SR-RBC is also an adaptive compensation method.

## 4. Simulations

In this section, the airborne radar is deployed with cylindrical arrays. The problem of range ambiguity is not considered and the configuration parameters are given in Table I. The spectrum estimations using different methods are first given to verify the advantages of sparse representation. Then the performance such as SCR improvement is tested.

Table I Configuration parameters

| Parameter | Symbol | Value |
| --- | --- | --- |
| Number of rings | $M$ | 4 |
| Number of arrays on each ring | $N$ | 4 |
| Number of pluses | $P$ | 16 |
| Platform velocity | $v$ | 300m/s |
| Pulse repetition interval | $PRI$ | 0.25ms |
| Range sample rate | $f_s$ | 5Mhz |
| Radar wavelength | $\lambda$ | 0.3m |
| Inter-ring spacing | $d$ | 0.15m |
| Radius of the ring | $r$ | 0.15m |
| Platform height | $H$ | 3000m |
| Clutter-to-noise ratio | $CNR$ | 30dB |

Figs.3 (a)-(d) give the actual clutter response, Fourier, RBC and SR-RBC spectral estimations respectively, where the clutter scenario is side-looking CFA. Due to insufficient space-time samples of the STAP processor, the Fourier spectrum has a high sidelobe and the resolution is limited. On the other hand, both RBC and SR-RBC can achieve high-resolution estimation. However, since RBC requires the temporal smoothing to obtain a CCM estimate with sufficient rank [10], the dimension of CCM estimate is reduced so that the RBC spectrum has a limited accuracy, i.e., the clutter power spread and some missing of the actual scatters along the clutter ridge. To avoid this, RBC makes the peak extraction in this sub-snapshot spectrum and then fits it with the mathematical model to estimate the configuration parameters [9-10]. However, this extraction is partly sensitive to the spurious peaks as well as other artifacts, which appear in the sub-snapshot spectrum. Moreover, even if the estimation of configuration parameters is obtained with high accuracy, the clutter power estimation using LS might lose some actual clutter scatters since the estimation is only carried out in the locations of the extracted peaks. Thus, the accuracy in the sub-snapshot spectrum limits the performance of RBC. On the other hand, as shown in Fig.3 (d), since SR-RBC can obtain a desirable sparse solution at each range cell using IRLS, there is no sub-snapshot tradeoff and high-accurate spectrum estimation is expected. Parallel simulation is carried out with non side-looking CFA in Figs.4, where similar conclusion is obtained. Next the SCR improvement by different STAP algorithms is given to illustrate the advantage of SR-RBC.

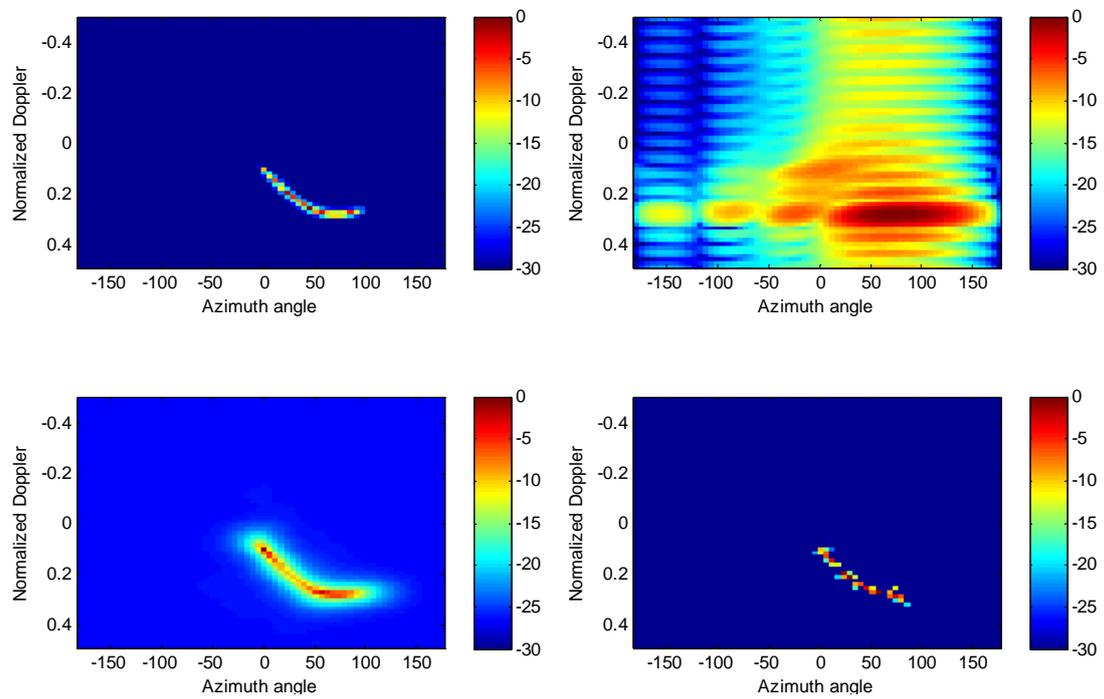

Figs.3 side-looking CFA (a) actual clutter response (b) Fourier estimation (c) RBC estimation (d) SR-RBC estimation

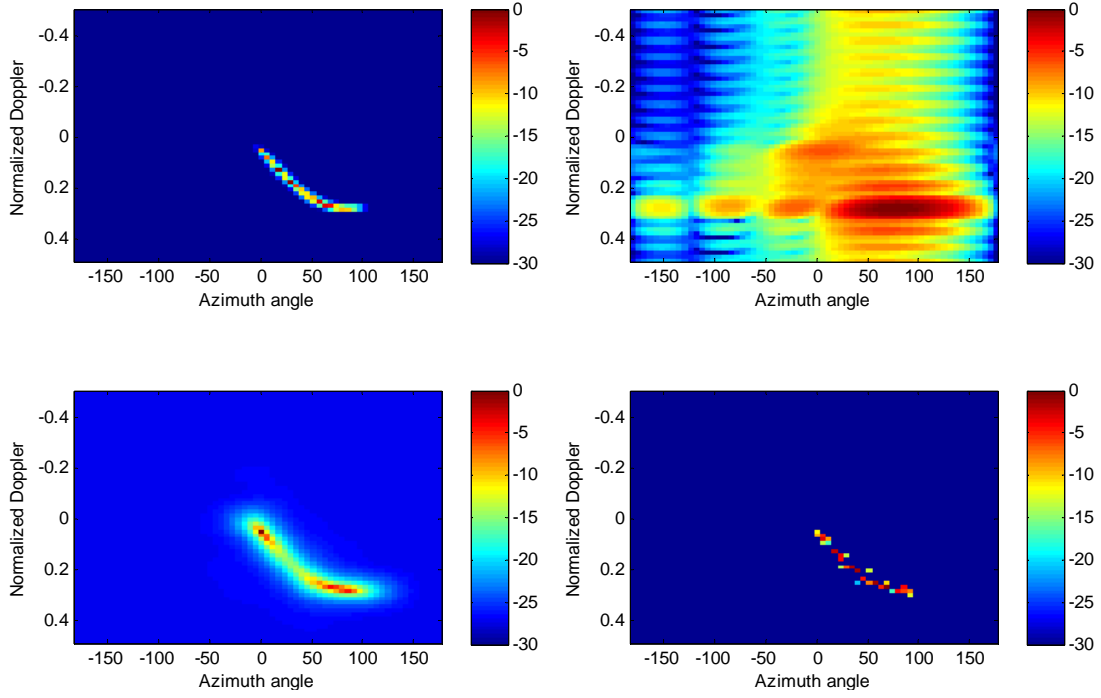

Figs.4 non side-looking CFA (a) actual clutter response (b) Fourier estimation (c) RBC estimation

(d) SR-RBC estimation

Conventionally, the efficiency of the STAP filter is evaluated by the normalized SCR improvement, which is defined as [1]

$$IF_{Loss} = \frac{SCR_{out}/SCR_{in}}{\left(SCR_{out}/SCR_{in}\right)_{opt}} = \frac{\left|\hat{\mathbf{w}}^H \mathbf{s}\right|^2}{\hat{\mathbf{w}}^H \mathbf{R}\hat{\mathbf{w}} \cdot \mathbf{s}^H \mathbf{R}^{-1}\mathbf{s}}, \qquad (28)$$

where the estimated adaptive filter is given as $\hat{\mathbf{w}} = \mu\hat{\mathbf{R}}^{-1}\mathbf{s}$, $\hat{\mathbf{R}}$ is the CCM estimation using a given technique (such as ADC, RBC or SR-RBC), $\mathbf{s}$ denotes the steering vector of the moving target, $\mathbf{R}$ is the actual CCM, and $tr(\mathbf{R})$ is the input clutter power. The slant range of the test cell is $R_s = 1.5H$, which is a typical short-range case. The training data is the adjacent range cells with the amount of $L = 40$. Fig.5 (a) gives the $IF_{Loss}$ performance of different STAP algorithms with side-looking CFA. Since the clutter behaves range-dependent, direct statistical method like LSMI has a degraded performance and the clutter notch is mismatch with the actual scenario. ADC aligns the peaks of the clutter ridge of the training data with that of the test cell. However, it only accomplishes partial compensation and the performance is still not desirable. On the other hand, RBC and SR-RBC make both mainlobe and sidelobe compensation. However, since the estimation of both the configuration parameters and clutter power distribution is based on the accuracy of the clutter spectrum, the RBC performance is limited. However, since SR-RBC can obtain high-accurate spectral estimation with full-snapshot, the corresponding training data after the transforming behaves more stationary, which brings desirable SCR improvement. Moreover, since SR-RBC directly estimate the Doppler response in the spectral domain, the estimation of configuration parameters such as velocity and crab angle are both avoided. Thus, as shown in Fig.5 (b), SR-RBC still preserves desirable performance with the non side-looking CFA.

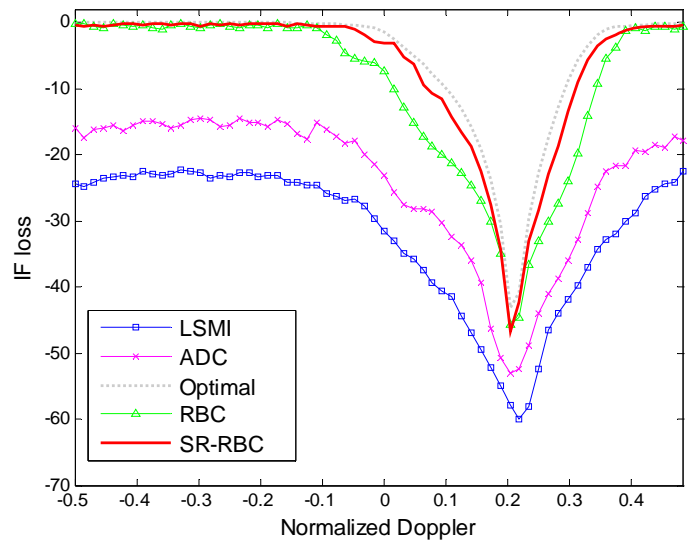

Figs.5 (a) IF Loss curves with side-looking CFA

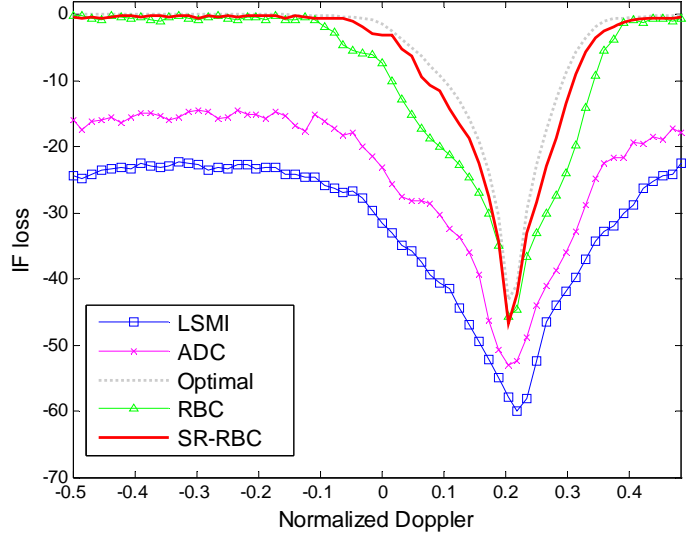

Figs.5 (b) IF Loss curves with non side-looking CFA

## 5. Conclusions

In this paper, we have analyzed the sparsity of the clutter spectral response in CFA and proposed a new compensation strategy called SR-RBC to deal with range-dependent clutter. The key advantage of SR-RBC is the capability of obtaining high-accurate spectral estimation with full-snapshot, which is owing to the technique of sparse representation. In this way, SR-RBC can acquire better transform matrix at each range cell so that the stationarity of the processed training data is further improved.

The following are some considerations for further research. First, the current overcomplete basis $\mathbf{\Phi}_k$ is fixed in sparse representation. However, due to the practical nonideal factors such as clutter internal motion and/or channel mismatch, this predefined overcomplete basis does not always match with the actual data and the corresponding sparsity might decrease. Therefore, solving the sparse representation problem where both overcomplete basis and actual sources are unknown seems to be quite important. Second, since IRLS is an iterative algorithm seeking to approximate the actual clutter spectrum, appropriate initial value and more adaptive mechanisms is necessary to approach the overall sparse solution.

## 6. Acknowledgment